\documentclass[a4paper,fleqn,usenatbib]{mnras}
\usepackage[T1]{fontenc}
\usepackage{ae,aecompl}
\usepackage{graphicx}	% Including figure files
\usepackage{amsmath}	% Advanced maths commands
\usepackage{amssymb}	% Extra maths symbols

\usepackage[utf8]{inputenc}
\usepackage{natbib}
\usepackage{epsfig}
\usepackage{appendix}
\usepackage{rotating}

\title[Detection of CH$_3$NCO in a solar-type protostar]
{Detection of methyl isocyanate (CH$_3$NCO) in a solar-type protostar}
\author[R. Mart\'in-Dom\'enech et al.]
{R. Mart\'in-Dom\'enech,$^{1}$ %\altaffilmark{1}, 
V. M. Rivilla,$^{2}$\thanks{E-mail: rivilla@arcetri.astro.it} %\altaffilmark{2}, 
I. Jim\'enez-Serra,$^{3}$ %\altaffilmark{3}, 
D. Qu\'enard,$^{3}$ %\altaffilmark{3}, 
\newauthor{L. Testi,$^{2,4,5}$, and J. Mart\'in-Pintado,$^{1}$} %\altaffilmark{1}}
\\
$^{1}$Centro de Astrobiolog\'ia (INTA-CSIC). Ctra de Ajalvir, Km. 4, Torrej\'on de Ardoz, 28850 Madrid, Spain\\
$^{2}$INAF/Osservatorio Astrofisico di Arcetri, Largo Enrico Fermi 5, I-50125, Florence, Italy\\
$^{3}$School of Physics and Astronomy, Queen Mary University of London, Mile End Road, London E1 4NS\\
$^{4}$ESO/European Southern Observatory, Karl Schwarzschild str. 2, D-85748, Garching, Germany\\
$^{5}$Excellence Cluster “Universe”, Boltzmann str. 2, D-85748 Garching bei Muenchen, Germany
}

\date{Accepted XXX. Received YYY; in original form ZZZ}

\pubyear{2017}

\begin{document}
\label{firstpage}
\pagerange{\pageref{firstpage}--\pageref{lastpage}}
\maketitle

% Abstract of the paper
\begin{abstract}
We report the detection of the prebiotic 
%complex organic 
molecule CH$_3$NCO in a solar-type protostar, IRAS16293-2422 B. 
%This species is one of the most abundant complex organic molecules tentatively detected 
A significant abundance of this species 
on the surface of the comet 67P/Churyumov-Gerasimenko 
has been proposed,
and 
%in the insterstellar medium 
it has 
%only 
recently 
been 
%found 
detected 
in hot cores around high-mass protostars. 
%We have used multi-frequency ALMA observations from 90 GHz to 265 GHz covering 
We observed IRAS16293-2422 B with ALMA in the 90 GHz to 265 GHz range, and detected 8 unblended transitions of CH$_3$NCO. 
%and 14 more transitions that appear blended with emission from other molecular species.
From our Local Thermodynamic Equilibrium analysis we derived an excitation temperature of 110$\pm$19 K and a column density of (4.0$\pm$0.3)$\times$10$^{15}$ cm$^{-2}$, which results in  an abundance of $\le$(1.4$\pm$0.1)$\times$10$^{-10}$ with respect to molecular hydrogen. 
This implies a CH$_3$NCO/HNCO and CH$_3$NCO/NH$_2$CHO column density ratios of $\sim$0.08. 
%The derived CH$_3$NCO/HNCO column density ratio is $\sim$0.08, of the same order of magnitude as those found in SgrB2(N) and Orion KL. However, the measured CH$_3$NCO/NH$_2$CHO ($\sim$0.08), although similar to that observed in SgrB2(N), it differs by factors 20-70 in Orion KL. 
Our modelling of the chemistry of CH$_3$NCO suggests that both ice surface and gas phase formation reactions of this molecule are needed to explain the observations. 
\end{abstract}


%%%%%%%%%%%%%%%%%%%%%%%%%%%%%%%%%%%%%%%%%%%%%%%%%%

%%%%%%%%%%%%%%%%% BODY OF PAPER %%%%%%%%%%%%%%%%%%

\section{Introduction}
\label{intro}

Understanding the origin of life is one of the main challenges of modern science. It is believed that some basic prebiotic chemistry could have developed in space, likely transferring prebiotic molecules to the solar nebula and later on to Earth. 
%Studies of the chemical composition of comets have indeed reported that these objects 
For example, comets exhibit a wide variety of complex organic molecules (or COMs) that are commonly detected in the ISM \citep[see, e.g.,][]{biver15}. 
Recently, the spacecraft Rosetta
%studied the chemical composition of the comet 67P/Churyumov-Gerasimenko and 
found evidence for the presence of several COMs of prebiotic interest on the 
%cometary 
surface 
of the comet 67P/Churyumov-Gerasimenko, 
using the COSAC mass spectrometer \citep[as e.g. glycoladehyde, CH$_2$OHCHO, or formamide, NH$_2$CHO;][]{goesmann15}, and in the coma of the comet using the ROSINA instrument \citep[with the detection of the amino acid glycine, and phosphorous;][]{altwegg16}. 
Among these species, the COSAC mass spectrometer 
%also proposed the detection of a new molecule, 
suggested the presence of 
methyl isocyanate (CH$_3$NCO) 
%in \textbf{the surface of} comet 67P/Churyumov-Gerasimenko 
with an abundance relatively high compared to other COMs \citep[][]{goesmann15}. CH$_3$NCO is the simplest isocyanate, which along NH$_2$CHO contains C, N, and O atoms,  
%Isocyanates are a family of complex organics that 
and could play a key role in the synthesis of amino acid chains known as peptides \citep{pascal05}. 
%%%%%%%%%%%%%%%%%%%%%%%%%%%%%%%%%%%%%%%%%%%%%%%%%%%%%%%%%
% SHOULD I INCLUDE THIS LATER IN THE DISCUSSION **** ??
%In particular, methyl isocyanate is the methyl derivative of isocyanic acid (HNCO), and its formation can be explained by UV photoprocessing of interstellar ice mantles \citep{goesmann15}. A fraction of these interstellar ice mantles is thermally desorbed to the gas phase in circumstellar environments around protostars (the so-called hot cores and hot corinos) in star-forming regions, while another fraction can be preserved in comets once the planetary system is formed.  
%%%%%%%%%%%%%%%%%%%%%%%%%%%%%%%%%%%%%%%%%%%%%%%%%%%%%%%%%
%After COSAC's tentative detection, 
CH$_3$NCO was subsequently detected in massive hot molecular cores such as SgrB2 N \citep{halfen15,belloche17} and Orion KL \citep{cernicharo16}. However, CH$_3$NCO remained to be reported in solar-type protostars. 
%\textbf{i.e., in a environment similar to the solar nebula where Earth formed.} 

IRAS 16293$-$2422 (hereafter IRAS16293) is located in the $\rho$ Ophiuchi cloud complex at a distance of 120 pc \citep{loinard08}, and it is considered an excellent template for astrochemical studies in low-mass protostars \citep[e.g.,][]{jorgensen11,jorgensen16,lykke16}. 
IRAS16293 is 
%a binary system with components 
composed by sources 
A and B, separated in the plane of the sky by $\sim$5$\arcsec$ ($\sim$600 AU), and whose masses are $\sim$0.5 M$_{\odot}$ \citep{looney00}. 
Their emission exhibits line profiles with linewidths of up to 8 km s$^{-1}$ for IRAS16293 A and $<$ 2 km s$^{-1}$ for IRAS16293 B. The narrow emission of IRAS16293 B, along with its rich COM chemistry, makes this object the perfect target to search for new COMs. 

%In this Letter, we report the first detection of CH$_3$NCO towards the solar-type protostar IRAS 16293$-$2422 B carried out with the Atacama Large Millimeter Array (ALMA). 
%\textbf{Detection of CH$_3$NCO towards the solar-type protostar IRAS 16293$-$2422 B carried out with the Atacama Large Millimeter Array (ALMA) is reported in this Letter and in \citet{ligterink17}, using transitions with low and high upper level energies, respectively.} 
In this letter we report the detection of CH$_3$NCO towards IRAS 16293-2422 B at frequencies $\le$250 GHz using the Atacama Large Millimeter Array (ALMA). Our results are consistent with those presented by \citet{ligterink17} using CH$_3$NCO transitions with frequencies $\ge$300 GHz.
%%%%%%%%%SOLO SI HAY ESPACIO EN LA LETTER%%%%%%%%%%%%%
%\textbf{The Letter is organised as follows: the observations are described in Sect. \ref{observations}, while the detection is reported in Sect. \ref{results},  
%and a chemical model reproducing the observed abundance for CH$_3$NCO is presented in Sect. \ref{model}.}

\section{Observations}
\label{observations}

%We have used multiple observations carried out toward IRAS16293-2422 with ALMA in Bands 3, 4, and 6. 
The analysis was carried out using the ALMA data from our own project (\#2013.1.00352.S), and all other public datasets in Bands 3, 4, and 6 available in the ALMA archive as of January 2017  (\#2011.0.00007.SV, \#2012.1.00712.S, and \#2013.1.00061.S). We note that we have excluded from our analysis other ALMA public datasets in Bands 7 and 8, i) to prevent any dust optical depth problems \citep[the dust continuum emission of IRAS16293 B is known to be very optically thick at frequencies $>$300 GHz, 
%which severely affects 
affecting
the line intensities of the molecular emission; see][]{zapata13,jorgensen16}; and ii) to limit the level of line confusion, which allows the correct subtraction of the continuum emission by selecting a suitable number of line-free channels in the observed spectra \citep[see e.g.][]{pineda12}.  All data matching our criteria were downloaded and re-calibrated using standard ALMA calibration scripts and the Common Astronomy Software Applications package\footnote{https://casa.nrao.edu}. The angular resolution of all datasets was sufficient to resolve source B from source A (with angular resolutions below 1.5$"$), and therefore the emission lines arising from source B are narrow with linewidths <2$\,$km$\,$s$^{-1}$. Continuum subtraction was performed in the uv-plane before imaging using line-free channels.

Our dataset covers a total bandwidth of $\sim$6 GHz split in 26 spectral windows spread between 89.5 and 266.5 GHz, with synthesised beam sizes ranging from 0.57$^{\prime\prime}\times$0.28$^{\prime\prime}$ to 1.42$^{\prime\prime}\times$1.23$^{\prime\prime}$. The velocity resolution falls between 0.14 and 0.30 km s$^{-1}$. For the analysis, a spectrum was extracted from each datacube using a circular support with size $\sim$1.6$^{\prime\prime}$ centered at the position of IRAS16293 B ($RA_{\rm J2000}$ = 16$h$ 32$m$ 22.61$s$, $DEC_{\rm J2000}$ = -24$^{\circ}$ 28$^{\prime}$ 32.44$^{\prime\prime}$). We note that the molecular emission from IRAS16293 B for the species considered in this study (e.g. CH$_3$NCO, NH$_2$CHO and HN$^{13}$CO), is compact and lies below 1.5$"$ \citep[see Figure 2 below and][]{coutens16}. Thus, although the ALMA datasets were obtained with different array configurations and UV coverage, we are confident that our extracted spectra contain all the emission from the hot corino and the analysed lines do not suffer from missing flux.

\section{Results}
\label{results}

\subsection{Detection of CH$_3$NCO}

\begin{figure*}
\centering
\includegraphics[scale=0.28]{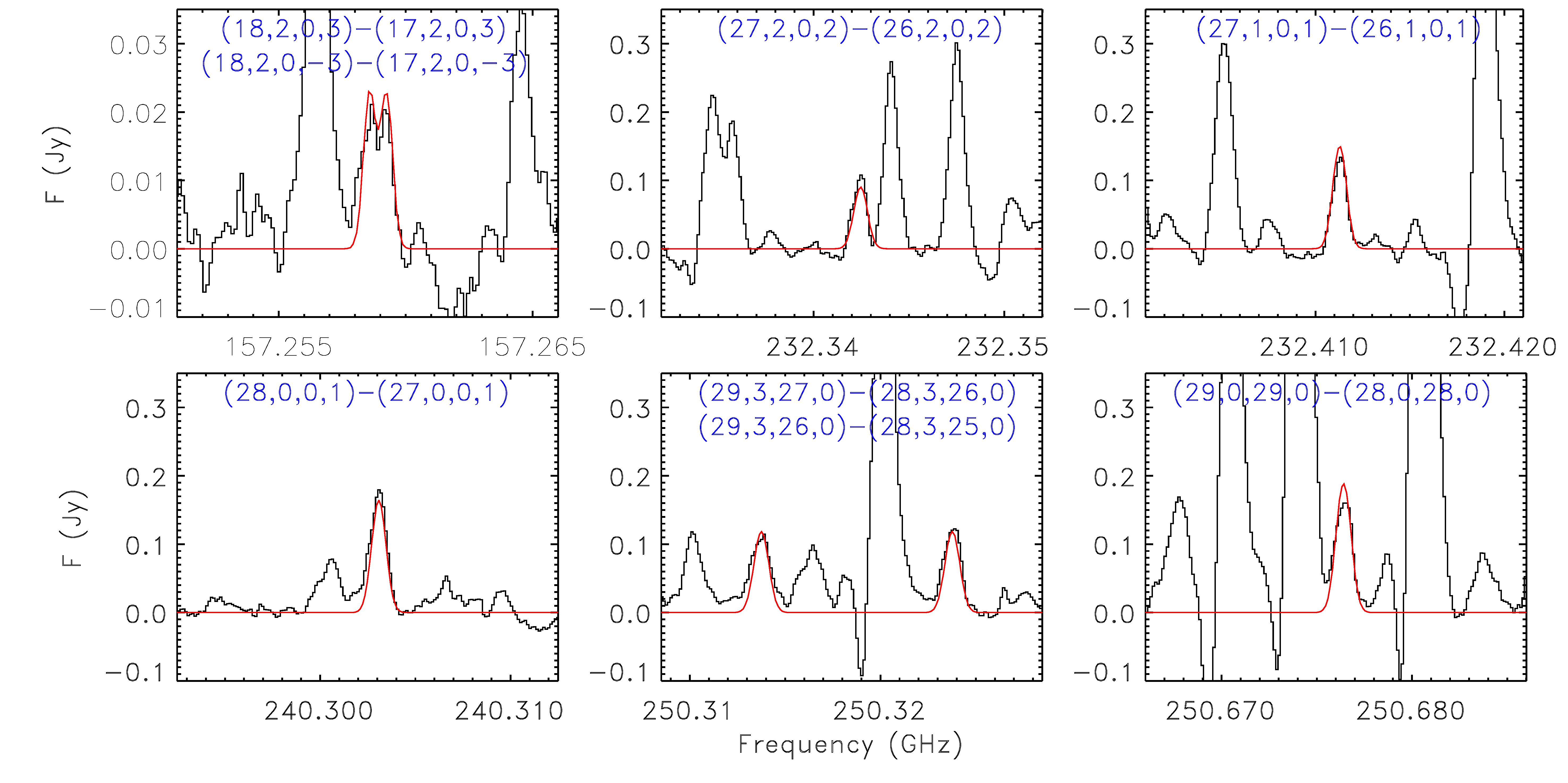}
\caption{CH$_3$NCO unblended lines measured toward IRAS16293 B with ALMA (solid black). 
Transitions are shown in every panel, while their rest frequencies are reported in Table \ref{table-unblended}. The synthetic LTE spectrum generated by MADCUBAIJ is overplotted in red.}
\label{figure-unblended}
\end{figure*}

\begin{figure}
\centering
\includegraphics[scale=0.5]{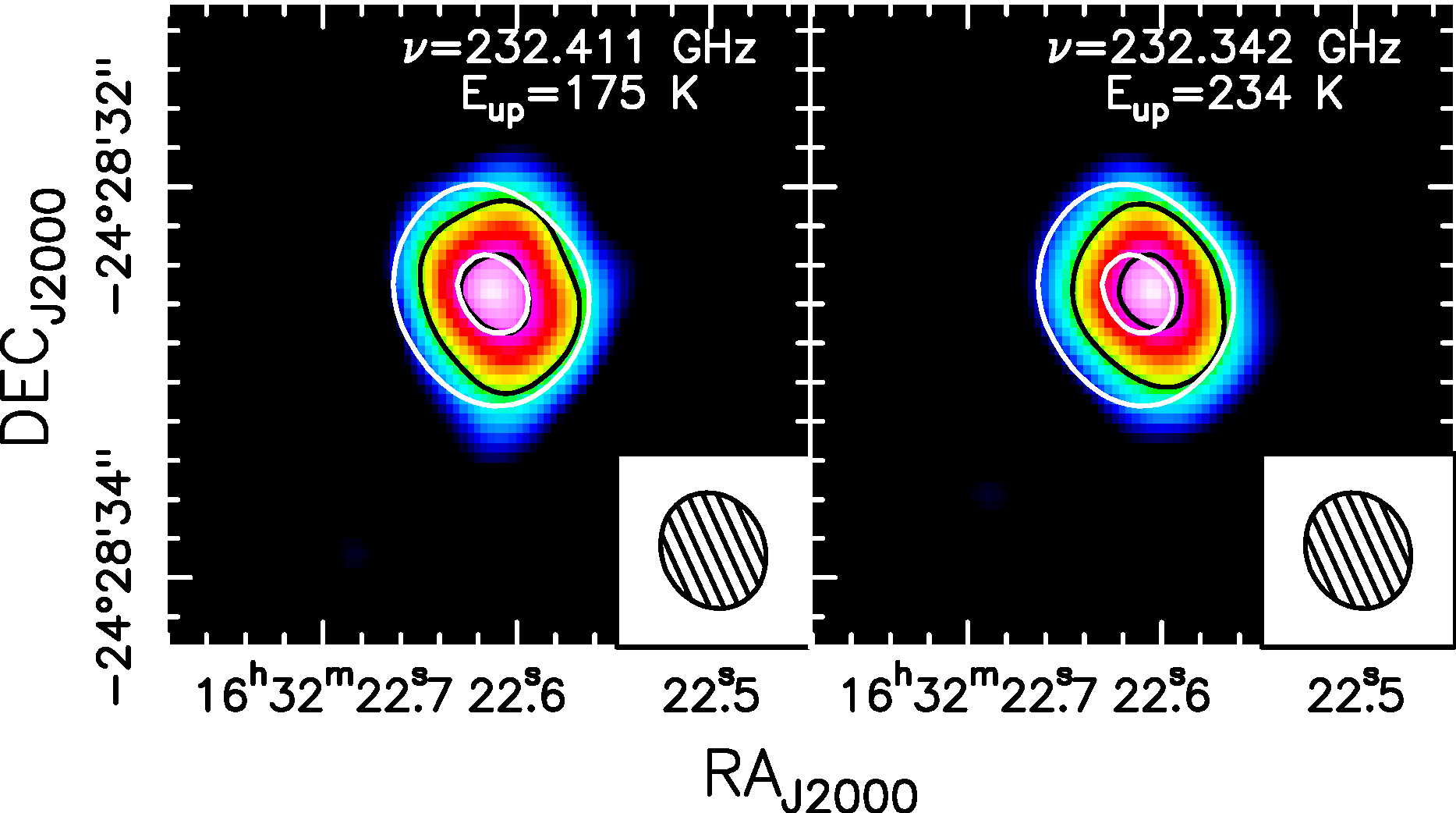}
\caption{Integrated intensity maps of two representative CH$_3$NCO unblended lines observed toward IRAS16293 B. Black contours indicate 50\% and 90\% of the peak line emission, while white contours indicate 20\%, and 80\% of the continuum peak emission at 232 GHz. 
The rest frequency and E$_{\rm up}$ of the transitions are shown in every panel (see also Table \ref{table-unblended}). Beam sizes are shown in the bottom right corner.}
\label{mapa}
\end{figure}

The rotational spectrum of CH$_3$NCO, with the A and E torsional states, has been studied by \citet{koput86} (from 8 to 40 GHz), and more recently by \citet{halfen15}  (from 68 to 105 GHz) and \citet[][from 40 to 363 GHz]{cernicharo16}. The identification of the lines was performed using the software MADCUBAIJ\footnote{Madrid Data Cube Analysis on ImageJ is a software developed in the Center of Astrobiology (Madrid, INTA-CSIC) to visualise and analyse single spectra and datacubes \citep{rivilla16a,rivilla16b}}, using the information from the Jet Propulsion Laboratory \citep[JPL;][]{pick98} and the Cologne Database for Molecular Spectroscopy spectral catalogs \citep[CDMS;][]{mull05}. 
We identified a total of 22 transitions of CH$_3$NCO, 8 out of which were unblended 
with upper level energies ranging from 175 to 233 K 
(see Table \ref{table-unblended}) using MADCUBAIJ. The remaining 14 lines appear contaminated by emission from other species.  
%This assures the unambiguous identification of the molecule at a 99$\%$-99.8$\%$ confidence level \citep{Halfen06}. 
The CH$_3$NCO lines peak at a radial velocity of $v_{LSR}$ = 2.7 km s$^{-1}$ and have linewidths of $\sim$1.1 km s$^{-1}$ (Fig. \ref{figure-unblended}), similar to those from other molecules in IRAS16293 B \citep[][]{jorgensen11}. 
\begin{table}
\tabcolsep 1.2pt
\begin{center}
\caption{Detected CH$_3$NCO unblended lines in IRAS16293 B.}
\label{table-unblended}
\begin{tabular}{c c c c c}
\hline
Frequency       & Transition	                & logA$_{ul}$	& E$_{\rm up}$  & Area 	           \\ % & Orion \\
(GHz)           & (J,K$_{a}$,K$_{c}$,m)         & (s$^{-1}$)& (K)           & Jy km s$^{-1}$   \\ % & \\
\hline
%139.631255	& (16,1,0,2)$-$(15,1,0,2)       & -4.0738       &      &  % & part. blend. \\
157.258419      & (18,2,0,3)$-$(17,2,0,3)       & -3.75       & 210      & 0.028 $\pm$ 0.009 \\% & \\
157.259087      & (18,2,0,-3)$-$(17,2,0,-3)     & -3.75       & 210      & 0.028 $\pm$ 0.009 \\% & \\
232.342227	& (27,2,0,2)$-$(26,2,0,2)	& -3.22	& 234	& 0.12 $\pm$ 0.04 \\% & Unbl.	\\
232.411044	& (27,1,0,1)$-$(26,1,0,1)	& -3.22	& 175	& 0.16 $\pm$ 0.06 \\% & Unbl.	\\
240.302835	& (28,0,0,1)$-$(27,0,0,1)	& -3.17	& 181	& 0.18 $\pm$ 0.06 \\% & Unbl.	\\
250.313498	& (29,3,27,0)$-$(28,3,26,0)	& -3.13	& 235	& 0.16 $\pm$ 0.05 \\% & Blend.	\\
250.323521	& (29,3,26,0)$-$(28,3,25,0)	& -3.13	& 235	& 0.16 $\pm$ 0.05 \\% & Blend.	\\
250.676140	& (29,0,29,0)$-$(28,0,28,0)	& -3.13	& 181	& 0.21 $\pm$ 0.07 \\% & Blend	\\
%344.650251      & (40,3,0,2)$-$(39,3,0,2)       & -3.35       & 448      & 0.24 $\pm$ 0.06 \\% & \\
%%346.358137     & (40,2,39)$-$(39,2,38)         & -3.2345       &      &  % & \\
%352.722522      & (41,-1,0,1)$-$(40,-1,0,1)     & -3.22       & 376      & 0.36 $\pm$ 0.09 \\% & \\
%353.216538      & (41,3,39,0)$-$(40,3,38,0)     & -3.27       & 465      & 0.31 $\pm$ 0.07 \\% & \\
%%353.345626     & (41,3,0)$-$(40,3,0)           & -3.3466       &      &  % & \\
\hline
\end{tabular}
\end{center}
\end{table}
MADCUBAIJ produces synthetic spectra assuming Local Thermodinamical Equilibrium (LTE) conditions.  
%taking into account the opacity of each transition and the source size. 
The comparison between the observed and the synthetic spectrum for the unblended ransitions can be used to derive the excitation temperature and total column density that best match the observations. 
%To this purpose we used the autofit tool of MADCUBAIJ. 
%We fixed the linewidth of the lines to 1.1 km s$^{-1}$, and the source size to 0.5$^{\prime\prime}$ for the emission originated in the compact warm and dense region around IRAS16932 B rich in COMs
%\citep[][]{jorgensen16,coutens16,lykke16}. Note that this source size is consistent with that measured from the continuum and molecular line emission (see Figure 2 and text below). 
We assumed a linewidth of 1.1 km s$^{-1}$, and the source size was constrained by the continuum emission to 0.5$^{\prime\prime}$ (see Figure \ref{mapa}),  
which agrees with the source size assumed in previous works \citep{jorgensen16,coutens16,lykke16}. 
The observed spectra and the corresponding LTE fitted synthetic spectrum for the 8 unblended lines detected are shown in Fig \ref{figure-unblended}. All CH$_3$NCO transitions were found to be optically thin ($\tau <$ 0.08), and thus our analysis is not affected by optical depth effects.
In addition, a careful check of the synthetic spectrum was performed to confirm that no dectectable transitions were missing from our observational data.
The detected CH$_3$NCO transitions 
%with energy levels in the 175$-$235 K range
are well reproduced by an excitation temperature of $T_{ex}$=110$\pm$19 K. This $T_{ex}$ is similar to that found for other COMs such as acetaldehyde or propanal in IRAS16293 B \citep[][]{lykke16}. The derived column density is $N$(CH$_{3}$NCO)=(4.0$\pm$0.3)$\times$10$^{15}$ cm$^{-2}$, 
which agrees with the column density reported in \citet{ligterink17} assuming the same source size and excitation temperature for the transitions with E$_{up}>$300 K detected at frequencies $\ge$320 GHz. 
The spatial distribution of CH$_3$NCO is shown in Fig. \ref{mapa} and it is coincident with the continuum emission. The measured deconvolved size is $\sim$0.5$"$, consistent with the assumed source size. %assumed in our LTE analysis. 

In order to estimate the CH$_3$NCO abundance, we have derived the H$_2$ column density by using the continuum flux measured at 232 GHz (1.4$\pm$0.05 Jy within a deconvolved size of 0.55$\arcsec\times$0.47$\arcsec$), and by assuming optically thin dust, a dust opacity of 0.009 cm$^{2}$ g$^{-1}$ \citep[thin ices in a H$_2$ density of 10$^{6}$ cm$^{-3}$; see][]{ossenkopf94} and a gas-to-dust mass ratio of 100. The estimated H$_2$ column density for T$_{dust}$=T$_{ex}$=110 K
%\footnote{At the high H$_2$ volume densities of IRAS16293 B, dust and gas are thermally coupled.} 
 (at these high densities, dust and gas are thermally coupled)
is $N$(H$_{2}$)=2.8 $\times$10$^{25}$ cm$^{-2}$, consistent with that estimated by \citet{jorgensen16} at higher frequencies. We however caution that this value should be considered as a lower limit since dust may be optically thick even at these low frequencies.  The derived abundance of CH$_{3}$NCO is $\chi$(CH$_{3}$NCO)=(1.4$\pm$0.1)$\times$10$^{-10}$ and it should be considered as an upper limit. 

Transitions corresponding to two isomers of methyl isocyanate, CH$_3$CNO and CH$_3$OCN, were not detected in our dataset, 
and 3$\sigma$ upper limits of 2.7$\times$10$^{13}$ cm$^{-2}$, and  5.1$\times$10$^{14}$ cm$^{-2}$ (respectively) were extracted 
assuming the same linewidth and excitation temperature. 
This upper limits lead to column density ratios of CH$_3$NCO/CH$_3$CNO$\ge$150 and CH$_3$NCO/CH$_3$OCN$\ge$8.

%The CH$_3$NCO emission is thus originated in an unresolved compact region around the protostar in IRAS16293 B. 

\subsection{Chemically-related species: HNCO and NH$_2$CHO}

In Orion KL, CH$_3$NCO shows the same spatial distribution as HNCO and NH$_2$CHO 
\citep{cernicharo16} and therefore they are thought to be chemically related. Several transitions of HNCO and NH$_2$CHO, and of some of their isotopologues, are also covered and detected in our dataset. The HNCO and NH$_2$CHO lines are optically thick \citep{coutens16} and their column densities have been inferred using the HNC$^{18}$O and NH$_2^{13}$CHO isotopologues. 
Five unresolved transitions of HNC$^{18}$O are found at 250 GHz with E$_{up}$=122 K. 
%By using MADCUBAIJ and 
For 
a fixed excitation temperature of $T_{ex}$=110 K, 
(the $T_{ex}$ derived for CH$_3$NCO; see Section 3.1) 
we obtain a column density of $N$(HNC$^{18}$O)=(9.7$\pm$3.8)$\times$10$^{13}$ cm$^{-2}$. By assuming an isotopic ratio $^{16}$O/$^{18}$O = 500 (Wilson \& Rood 1994), the derived total column density of HNCO is $N$(HNCO)=(4.9$\pm$1.9)$\times$10$^{16}$ cm$^{-2}$, which yields an abundance of (1.8$\pm$0.7)$\times$10$^{-9}$. 

For NH$_2^{13}$CHO, three lines are detected at 156.957 GHz, 157.097 GHz, and 239.628 GHz, 
with E$_{up}$ = 58$-$98 K. 
Their emission is fitted 
%in MADCUBAIJ by using an 
with an excitation temperature of $T_{ex}$=75 K
%\textbf{(due to the lower values of E$_{up}$ covered by the NH$_2^{13}$CHO lines)}, 
and a column density of $N$(NH$_2^{13}$CHO)=(7.6$\pm$3.7)$\times$10$^{14}$ cm$^{-2}$. 
The derived $T_{ex}$ is slightly lower than that obtained for CH$_3$NCO, possibly due to the lower values of E$_{up}$ covered by the NH$_2^{13}$CHO lines compared to those of CH$_3$NCO. We note however, that both species show the same spatial extent \citep[see Figure 2 and][]{coutens16} and therefore, they likely trace the same gas.  
By assuming an isotopic ratio $^{12}$C/$^{13}$C=68 \citep{milam05}, the derived total column density is $N$(NH$_2$CHO)=(5.2$\pm$2.5)$\times$10$^{16}$ cm$^{-2}$, which gives an abundance of (1.9$\pm$0.9)$\times$10$^{-9}$. As for CH$_3$NCO, these abundances should be considered as upper limits.

\subsection{Comparison with other sources}

%The rotational temperature found for CH$_3$NCO in IRAS16293 B (110$\pm$19 K) is slightly lower than that derived in Orion KL ($\sim$150 K) and SgrB2(N) \citep[$\sim$200 K;][]{cernicharo16}.
The abundance of (1.4$\pm$0.1)$\times$10$^{-10}$ measured for CH$_3$NCO toward IRAS16293 B  is similar to that found in SgrB2(N) \citep[1.7$\times$10$^{-9}$ and 1.0$\times$10$^{-9}$ for the two $V_{LSR}$ components; see][]{cernicharo16}. In Table \ref{comparacion}, we present the comparison between the abundance ratios CH$_3$NCO/HNCO and CH$_3$NCO/NH$_2$CHO measured in IRAS16293 B with those from the three sources where CH$_3$NCO has also been detected \citep[e.g. SgrB2(N), Orion KL, and 67P/Churyumov-Gerasimenko;][]{goesmann15,halfen15,belloche17,cernicharo16}. 
Since we have estimated the column densities considering the same emitting region, the derived ratios are likely independent on the assumed source size and the derived H$_2$ column density. %toward IRAS16293 B.

From Table \ref{comparacion}, we find that the CH$_3$NCO/HNCO column density ratio in IRAS16293 B is of the same order as those measured in SgrB2(N) and Orion KL. However, it is a factor of $\sim$50 lower than in comet 67P/Churyumov-Gerasimenko. 
%This may suggest a significant production of CH$_3$NCO after the warm hot corino phase by further UV photoprocessing of HNCO-containing ice mantles in the solar nebula \citep[][]{goesmann15}. 
We note however that the COSAC detections are tentative and therefore the abundance ratios in column 7 of Table \ref{comparacion} should be taken with caution.
The CH$_3$NCO/NH$_2$CHO column density ratio in IRAS16293 B is similar to that observed in SgrB2(N), while it is factors 20-70 lower than those measured in Orion KL, and a factor of 10 lower than that in comet 67P/Churyumov-Gerasimenko. In Section 4, we explore the formation routes for CH$_3$NCO and compare the measured ratios with those predicted by chemical modelling.

\section{Chemical modelling}
\label{model}

For the chemical modelling of CH$_3$NCO, HNCO, and NH$_2$CHO in IRAS16293 B, we have used the gas-grain chemical code UCL\_CHEM recently re-written by Holdship et al., (submitted)\footnote{UCL\_CHEM can be downloaded at https://uclchem.github.io/.}. UCL\_CHEM's chemical network contains 310 species (100 of them are also on the grain surface) and 3097 reactions. Gas phase reactions are taken from the UMIST database \citep{mcelroy2013}, while dust grain surface processes include thermal desorption \citep[as in][]{viti2004} and non-thermal desorption processes such as direct UV desorption, cosmic ray-induced UV photons desorption, direct cosmic ray desorption, and H$_2$ formation mechanism desorption \citep[see][]{roberts2007}. Recently, diffusion following the formalism from \citet{hasegawa1992}, and chemical reactive desorption \citep[following the experimentally-derived formula of][]{minissale2016} have also been included in UCL\_CHEM (Qu\'enard et al. in prep.).

To model the chemistry of CH$_3$NCO, HNCO and NH$_2$CHO, we expanded UCL\_CHEM's chemical network by including recently proposed gas phase and grain surface formation routes. For the gas phase formation of CH$_3$NCO, \citet{halfen15} proposed the following reactions:

\begin{eqnarray}
	\mathrm{HNCO/HOCN + CH_3} &\longrightarrow& \mathrm{CH_3NCO + H}\\
	\mathrm{HNCO/HOCN + CH_5^+} &\longrightarrow& \mathrm{CH_3NCOH^+ + H_2}\\
	\mathrm{CH_3NCOH^+ + e^-} &\longrightarrow& \mathrm{CH_3NCO + H}
\end{eqnarray}

\noindent
Note that no gas-phase destruction route was proposed in their study. We have also included the reaction
\begin{eqnarray}
	\mathrm{CH_3NCOH^+ + e^-} &\longrightarrow& \mathrm{CH_3 + HOCN},
\end{eqnarray}
\noindent
to account for the fact that CH$_3$NCOH$^+$ may fragment into smaller products. For the grain surface formation, \citet{belloche17} and \citet{cernicharo16} proposed that CH$_3$NCO could be formed through the grain surface reactions:
\begin{eqnarray}
	\mathrm{CH_3 + OCN} &\longrightarrow& \mathrm{CH_3NCO}\\
	\mathrm{CH_3 + HNCO} &\longrightarrow& \mathrm{CH_3NCO + H}.
\end{eqnarray}
These reactions have been found to be efficient experimentally \citep{ligterink17}. Furthermore, one of the possible formation routes of N-methylformamide (N-CH$_3$NHCHO) may involve successive addition of hydrogen atoms to CH$_3$NCO:
\begin{eqnarray}
	\mathrm{CH_3NCO + H} &\longrightarrow& \mathrm{CH_3NHCO}\label{hadd1}\\
	\mathrm{CH_3NHCO + H} &\longrightarrow& \mathrm{CH_3NHCHO}.
\end{eqnarray}
\noindent
where reaction (\ref{hadd1}) has an activation energy of $\sim$2500\,K \citep{belloche17}. The HNCO network (both in the gas and on the grain surface), together with that of its isomers HCNO and HOCN \citep[see][]{quan2010}, have also been included in UCL\_CHEM.

The physical conditions and the chemical composition of the IRAS16293 B hot corino were modelled using a three phase model. In Phase 0, we followed the evolution of the chemistry in a diffuse cloud (size of $\sim$0.6\,pc and $A_V=2$\,mag) by assuming a
constant density of $n_\textrm{H}=10^{2}$\,cm$^{-3}$ and a temperature of $T_{kin}=10$\,K for $\sim$10$^6$ yrs. We assume an interstellar radiation field of $G_0=1$ Habing and the standard cosmic ray ionisation rate of $1.3\times10^{-17}$\,s$^{-1}$. The elemental abundances considered in this model are taken from \citet[][model EA1]{wakelam2008}. 

In Phase 1 of our model, we follow the chemistry during the pre-stellar core phase assuming a constant temperature of $T_{kin}=10$\,K while we increase the core's gas density from 10$^{2}$$\,$cm$^{-3}$ to $5\times10^{8}$\,cm$^{-3}$ (this value is consistent with that measured in IRAS16293 B). In Phase 2, the chemical evolution of the hot corino is modelled by assuming a constant H$_2$ gas density ($5\times10^{8}$\,cm$^{-3}$) while gradually increasing the gas temperature from 10$\,$K to 110$\,$K during the first 10$^5$ yrs. After then, the temperature is kept constant and the chemistry is followed during $10^6$\,yrs. 

The best fit to our observations is found for a dynamical age of $\times$10$^4$ yrs, with a predicted abundance for CH$_3$NCO of [CH$_3$NCO]=$6.0\times10^{-10}$ (with respect to H$_2$), i.e. a factor of 4 higher than that measured in IRAS16293 B (1.4$\times$10$^{-10}$; Section 3.1). We note that these time-scales are consistent with those estimated for this source \citep[see, e.g.,][]{bottinelli14,majumdar16}. In this model, HNCO (the parent molecule of CH$_3$NCO) is formed on the surface of dust grains; and once the temperature reaches $\sim$100 K, HNCO is thermally desorbed and incorporated into the gas phase, allowing the gas-phase formation of CH$_3$NCO to proceed \citep[see reactions above and][]{halfen15}. We note that CH$_3$NCO is also formed on grain surfaces in our model. However, this mechanism by its own is not sufficient to explain the observed abundances of this molecule in IRAS16293 B. Therefore, our modelling shows that formation both in the ices and in the gas phase is required to explain the observed abundance of CH$_3$NCO in IRAS16293 B. 
%CH$_3$NCO is indeed produced efficiently on grain surfaces as proposed by \citet{ligterink17}, but this molecule is subsequently destroyed on dust  grains via reaction (8) above.} 
We note that, while the HNCO abundance predicted by our model (3.8$\times$10$^{-9}$) also agrees well with that observed in IRAS16293 B (1.8$\times$10$^{-9}$), the abundance of NH$_2$CHO is underproduced by a factor of 10. As a result, while the CH$_3$NCO/HNCO abundance ratio is well reproduced by our model, the CH$_3$NCO/NH$_2$CHO ratio differs from that observed by a factor of $\sim$40 (see Table \ref{comparacion}). 
We also note that our model perfectly reproduces the upper limits of CH$_3$NCO measured in cold cores such as L1544 \citep[$\leq$2-6$\times$10$^{-12}$;][Qu\'enard et al., in prep]{izaskun16}.

\begin{table}
\tabcolsep 1.2pt
\label{comparacion}
\caption[]{Comparison of the CH$_3$NCO/HNCO and CH$_3$NCO/NH$_2$CHO ratios measured in IRAS16293 B, SgrB2(N), Orion KL and comet 67P. Our modelling results of IRAS16293 B for T$_{gas}$=110$\,$K and time=4$\times$10$^4$ yrs}, are also shown.
\begin{center}
\begin{tabular}{ c c c c c c c}
\hline
           & \multicolumn{5}{c}{Protostars}&  Comet \\ \cline{2-6}
 Molecular & \multicolumn{2}{c}{Low-mass} & \multicolumn{3}{c}{High-mass}  &  \\ 

 ratio & \multicolumn{2}{c}{IRAS16293 B} & SgrB2(N) & \multicolumn{2}{c}{Orion KL}&67P\\ 
               &  obs. & model    &  &  A      &  B                & \\
\hline
CH$_3$NCO/HNCO      & 0.08 & 0.16 & 0.11 &0.02  &  0.06 & 4.33  \\
CH$_3$NCO/NH$_2$CHO & 0.08 & 3.53  & 0.06 &1.75  &  5.71 & 0.72  \\ 
%CH$_3$NCO/CH$_3$CN  & 0.21 & - & 0.10  &  0.04 & 4.33 \\
\hline
\end{tabular}
%\begin{footnotesieze}
 %{$^{a}$The detections on the surface of comet 67P/Churyumov-Gerasimenko are tentative.}
%\end{footnotesieze}

\end{center}
\end{table}

%To test the effects of a higher temperature on the chemistry of CH$_3$NCO, we also run a second model in which the temperature in the hot corino increases up to 250$\,$K. The predicted abundance of CH$_3$NCO 
%%increases slightly (4.8$\times$10$^{-10}$), 
%\textbf{is very similar}, which shows that its chemistry does not depend strongly on the maximum temperature in our model as long as it is high ($\ge$100 K) to induce ice sublimation.

We carried out an additional test including the isomers of CH$_3$NCO, for which their upper limits have been measured (see Section$\,$\ref{results}). We have assumed that CH$_3$OCN and CH$_3$CNO experience the same reactions as CH$_3$NCO at the same rates, although this assumption is highly uncertain given the lack of experimental data. The abundance of CH$_3$NCO changes only by a factor of 1.1, but CH$_3$OCN and CH$_3$CNO are overproduced by factors $\geq$10-100. This means that their associated reaction rates need to be lowered by several orders of magnitude to match the observed upper limits. The full chemical network of CH$_3$NCO and its isomers will be discussed in detail in Qu\'enard et al. (in prep.).

%In summary, we present the first detection of CH$_3$NCO toward a solar-type protostar. The derived abundance of 1.4$\times$10$^{-10}$ is similar to that measured in hot cores, and the observed CH$_3$NCO/HNCO ratio is similar to that inferred in SgrB2(N) and Orion KL. The ratio with other chemically related N-, C-, O-bearing species suchs as NH$_2$CHO shows larger discrepancies across different objects, although it agrees well with that derived in SgrB2(N). The observed CH$_3$NCO abundance in IRAS16293 B is well reproduced by a chemical model that considers both gas-phase and grain surface reactions, {\bf which suggests that both formation routes are needed to explain the observations.}

\section*{Acknowledgments}
This Letter makes use of the following  ALMA data: ADS/JAO.ALMA\#2011.0.00007.SV, 
%ADS/JAO.ALMA
\#2012.1.00712.S, 
%ADS/JAO.ALMA
\#2013.1.00061.S, and 
%ADS/JAO.ALMA
\#2013.1.00352.S. 
ALMA is a partnership of ESO (representing its member states), NSF (USA) and NINS (Japan), together with NRC (Canada), NSC and ASIAA (Taiwan), and         KASI         (Republic
of          Korea),          in          co-operation          with          the          Republic          of          Chile.                The          Joint          ALMA          Observatory          is
operated        by        ESO,        AUI/NRAO        and        NAOJ
This research was partially financed by the Spanish MINECO under project AYA2014-60585-P, by the  Italian  Ministero  dell’
Istruzione, Università e Ricerca, through the grant Progetti Premiali 2012 – iALMA (CUP
C52I13000140001), and by the
Gothenburg Centre  for  Advanced  Studies  in  Science  and  Technology, where the re-calibration and re-imaging of all the ALMA Archive data on IRAS 16293-2422 was carried out  as  part  of  the  GoCAS program
“Origins of Habitable Planets”.  
R.M.-D. benefited from a FPI grant from Spanish MINECO. 
I.J.-S. acknowledges the financial support from an STFC Ernest Rutherford Fellowship and Grant (projects ST/L004801/2 and ST/M004139/2). 
J.M.-P. acknowledges partial support by the MINECO under grants FIS2012-39162-C06-01, ESP2013-47809-C03-01, and ESP2015-65597-C4-1. 
We thank an anonymous referee, and Dr. Wing Fai-Thi for providing useful comments on the manuscript.

\bsp	% typesetting comment
\label{lastpage}
\end{document}